\documentclass[prl,aps,twocolumn,showpacs]{revtex4}
 \newcommand{\mytitle}[1]{
 \twocolumn[\hsize\textwidth\columnwidth\hsize
 \csname@twocolumnfalse\endcsname #1 \vspace{1mm}]}
 \newcommand{\beq}{\begin{equation}}
 \newcommand{\eeq}{\end{equation}}
 \newcommand{\bea}{\begin{eqnarray}}
 \newcommand{\eea}{\end{eqnarray}}

\usepackage{graphics}
\usepackage{graphicx}
\usepackage{amssymb}
\usepackage{color}
\usepackage{bm}

\begin{document}

\title{Quantum criticality of the two-channel 
pseudogap Anderson model:  
Universal scaling in linear and non-linear conductance}
\author{Tsan-Pei Wu$^{1}$ and Chung-Hou Chung$^{1,2}$}
\affiliation{
$^{1}$Electrophysics Department, National Chiao-Tung University,
HsinChu, Taiwan, 300, R.O.C. \\
$^{2}$National Center for Theoretical Sciences, HsinChu, Taiwan, 300, R.O.C. \\
}

\date{\today}

\begin{abstract} 
The quantum criticality of the two-lead two-channel pseudogap 
Anderson model is studied. 
Based on the non-crossing approximation,
we calculate both the linear and nonlinear conductance of the model 
at finite temperatures with a voltage bias and a power-law vanishing   
conduction electron 
 density of states, $\rho_c(\omega)\propto |\omega-\mu_F|^r$ ($0<r<1$) near the Fermi energy $\mu_F$. 
Equilibrium and non-equilibrium quantum critical properties 
at the two-channel Kondo (2CK) 
to local moment (LM) phase transition are addressed by   
extracting universal scaling 
functions in both linear and non-linear    
conductances, respectively. Clear distinctions are found on 
the critical exponents between linear and non-linear conductance. 
The implications of these two distinct quantum critical properties for 
the non-equilibrium quantum criticality in general are 
discussed.    
\end{abstract}
\pacs{}

\maketitle
\section{\bf I. Introduction.}

\begin{figure}[t]
\begin{center}
\includegraphics[angle=0,width=1.0 \linewidth]{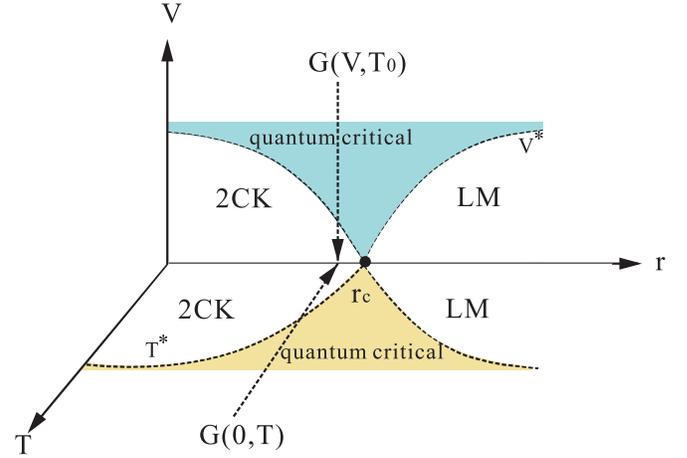}
\end{center}
\par
\vspace{-0.0cm}
\caption{Schematic 3D phase diagram of 2CK-LM quantum phase transition of 
the two-lead 2-channel pseudogap Anderson model as functions of voltage bias $V$, temperature $T$ and power-law 
exponent $r$ of the pseudogap fermion bath.
}
\label{phase}
\end{figure} 

Quantum phase transitions (QPTs)\cite{QPT}, the zero-temperature phase  
transitions due to quantum fluctuations, are of fundamental importance 
in condensed matter systems. Near the transitions, 
these systems show non-Fermi liquid behaviors manifested 
in universal power-law 
scaling in all thermodynamic observables.  
Recently, QPTs in quantum impurity problems\cite{vojta-review}, 
have attracted 
much attention recently due to their relevance for the  
nano-systems, such as: quantum dots\cite{QD}, realized experimentally. 
The well-known 
Kondo effect\cite{kondo}, the antiferromagnetic spin correlations 
between impurity and conduction electrons, plays a crucial role 
in understanding their low temperature behaviors.  
New scaling laws are expected to occur when the QPTs are associated 
with the Kondo breakdown in these systems either in equilibrium 
(at finite temperatures) or under non-equilibrium conditions 
(at finite voltage bias). 
Of particular interest 
lies in QPTs out of equilibrium\cite{QPT-noneq} where distinct 
universal scalings are expected 
in contrast to the counterparts in equilibrium. 

A fascinating playground to address this issue 
is the exotic two-channel Kondo (2CK)\cite{2CK,Affleck-Ludwig,kroha,Fendley,nayak,sela} systems 
with non-Fermi liquid ground state due to overscreening of $s=1/2$ 
impurity spin by two independent conduction reservoirs. 
Much of the theoretical effort has been made   
for the 2CK physics, including:    
via Bethe ansatz~\cite{Bethe}, conformal field theory~\cite{CFT}, 
bosonization~\cite{bosonization} and the 
numerical renormalization group~\cite{2CK-NRG}. 
Experimentally, the 2CK ground state has been realized in  
semiconductor quantum dots~\cite{Goldhaber},  
magnetically doped nanowires, and metallic glasses~\cite{Cox_a,Cichorek}. 
Recently, Kondo physics in magnetically doped graphene has attracted much 
attention for the possible 2CK physics as well as 
the pseudogap local density of states (LDOS) $\rho_c(\omega)$ 
which vanishes linearly due to the Dirac spectrum: 
$\rho_c(\omega) \propto |\omega|^r$ with $r=1$~\cite{sengupta,Vojta-2,Mattos}. 
This leads to QPT from 
the Kondo screened phase to the 
unscreened local moment (LM) phase with decreasing Kondo correlation 
due to insufficient DOS of conduction electrons. 
In fact, such transitions exist in the more general framework 
of pseudogap 
Kondo (or Anderson) models with power-law exponent $0<r<1$, which 
have been extensively studied~\cite{Fradkin,Gonzalez-Buxton.98, Ingersent,Glossop_a,Vojta4}. 
However, relatively less is known on the more exotic 2CK-LM QPT in the 
pseudogap Anderson (Kondo) models~\cite{Vojta3,Zamani.12}, 
in particular, when the system is subject to a 
non-equilibrium condition\cite{Zamani.13,QPT-noneq}. 

Motivated by these developments, we address the 
2CK-LM quantum phase transition  
in the two-channel pseudogap Anderson model both 
in and out of equilibrium in a Kondo quantum dot subject 
to a voltage bias and finite temperature. 
By studying experimentally accessible steady-state transport, we search for 
universal scalings in both linear and non-linear conductance near criticality 
via a large-N method based on Non-Crossing Approximation (NCA)~\cite{Cox_b,Meir,kroha,vojta-largeN}, a reliable approach for multi-channel Kondo systems with 
non-Fermi liquid ground states. 
A fundamentally important but less addressed 
issue--non-equilibrium quantum criticality-- is emphasized here 
by identifying and comparing different universal 
scaling behaviors between equilibrium (zero bias) and non-equilibrium 
(finite bias) conductances near the 2CK-LM transition\cite{Wu-TP}.

\section{\bf II. The Model Hamiltonian.} 

\begin{figure}[t]
\begin{center}
\includegraphics[width=1.0\linewidth]{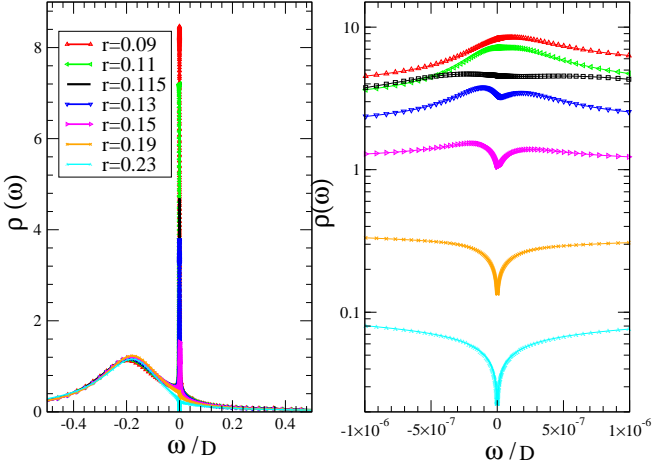}
\end{center}
\par
\vspace{-0.5cm}
\caption{The equilibrium 
impurity spectral function $\rho(\omega)$ (in arbitrary unit) 
versus energy $\omega$ (in units of the bandwidth $D$) with  
$V=0$, $\Gamma=0.28$, $\epsilon_d=-0.2$, $T=5\times10^{-7}$. 
(a). The spectral function $\rho(\omega)$ in the full energy range. 
For $r<r_c$ ($r_c\sim 0.115$), the Kondo peaks are located 
at $\omega=0$ (Fermi energy), while for $r>r_c$, dips near $\omega=0$ are 
developed. (b). Same plots as in (a) with energies close to the peaks/dips.}
\vspace{0.5cm}
\label{DOSfig}
\end{figure}
The Hamiltonian of the two-lead two-channel single-impurity pseudogap 
Anderson model is formulated within the Non-Crossing Approximation 
(NCA)~\cite{Cox_b,Meir,kroha,vojta-largeN}, 
a large-N approach based on the $SU(N)\times SU(M)$ generalization of the 
$SU(2)$ model with $N\rightarrow \infty,M\rightarrow \infty$ 
being the number of degenerate flavors of spins 
$\sigma=\uparrow,\downarrow,\cdots N$ and the number of Kondo screening 
channels $K=1,2,\cdots M$. In the physical $SU(2)$ two-channel Kondo system, 
$N=M=2$. The two leads are described by
a  power-law vanishing density of states (DOS) at Fermi energy 
defined as $\rho_{c}(\omega)\sim|\omega|^r\Theta(D-|\omega|)$ with $0<r<1$, 
where $D=1$ is the bandwidth cutoff. Graphene and high$T_c$-cuprate 
superconductors are possible realizations of the pseudogap leads with $r=1$, 
while semiconductors with soft gaps are candidates with $0<r<1$. 
The Hamiltonian reads~\cite{Meir,kroha}
\begin{eqnarray}
H&=&\sum_{\tau, \alpha, \sigma, k}(\epsilon_{k\sigma}-\mu_\alpha )c^{\alpha\dag}_{k\sigma\tau}c^{\alpha}_{k\sigma\tau}
+\sum_{\sigma,\tau}\epsilon_{\sigma}d^{\dag}_{\sigma,\tau}d_{\sigma,\tau}\nonumber \\
&+&
\frac{1}{2}U \sum_{\sigma,\sigma^\prime,\tau,\tau^\prime} n_{\sigma,\tau}n_{\sigma^{'},\tau^\prime} (1-\delta_{\sigma,\sigma^\prime} \delta_{\tau,\tau^\prime})\nonumber \\
&+& \sum_{\tau, \alpha, \sigma, k}(V^{\alpha}_{k\sigma}c^{\alpha \dag}_{k\sigma\tau}d_{\sigma}+h.c.)\nonumber \\
\label{Hamiltonian}
\end{eqnarray}
where $\mu_{\alpha=L/R}= \pm V/2$ 
is the chemical 
potential of lead $\alpha=L/R$. 
The operators $c^{\alpha\dag}_{k\sigma\tau}$ ($c^{\alpha}_{k\sigma\tau}$) 
create (destroy) an electron in the leads with momentum $k$. 
Spin flavors are represented by $\sigma,\sigma^\prime=1,\cdots N$ and $\tau,\tau^\prime=1,\cdots M$ 
corresponds to $M$ independent electron reservoirs. Here, $N=M=2$. 
The $U$ term describes the on-site Coulomb energy on the impurity, and 
 $V^\alpha_{k\sigma}$ 
represents the hybridization strength between the graphene electrons
and the impurity. 

The Hamiltonian Eq.~(\ref{Hamiltonian}) can be solved via NCA in the 
large-U limit $U\rightarrow \infty$ with the following Hamiltonian 
in the pseudofermion slave-boson representation: 
\begin{eqnarray}
H &=&\sum _{k \sigma\tau \alpha} (\epsilon_k-\mu_{\alpha}) 
c^{\alpha \dag}_{k \sigma\tau}c^{\alpha}_{k \sigma \tau}
+\epsilon_d \sum_\sigma f^{\dag}_\sigma f_\sigma \nonumber \\
&+&\sum_{k \sigma\tau\alpha}(V^{\alpha}_{k \sigma} (f^{\dag}_{\sigma} b_{\tau}c^{\alpha}_{k \sigma\tau})+h.c)\nonumber \\
\label{large-U-H}
\end{eqnarray}
where 
the local (impurity) electron operator $d^{\dagger}_{\sigma,\tau}$ 
is decomposed in the pseudofermion representation 
as a product of pseudofermion $f^\dagger_\sigma$ and a slave-boson 
$b_\tau$: $d^{\dagger}_{\sigma,\tau}=f^{\dagger}_{\sigma}b_{\tau}$ subject 
to the local 
 constraint $Q=\sum_{\sigma}f^{\dagger}_{\sigma}f_
{\sigma}+\sum_{\tau}b^{\dagger}_{\tau}b_{\tau}=1$ via the Lagrange multiplier 
$\lambda$ to ensure single 
occupancy on impurity. 
\begin{figure}[t]
\begin{center}
\includegraphics[width=0.9\linewidth]{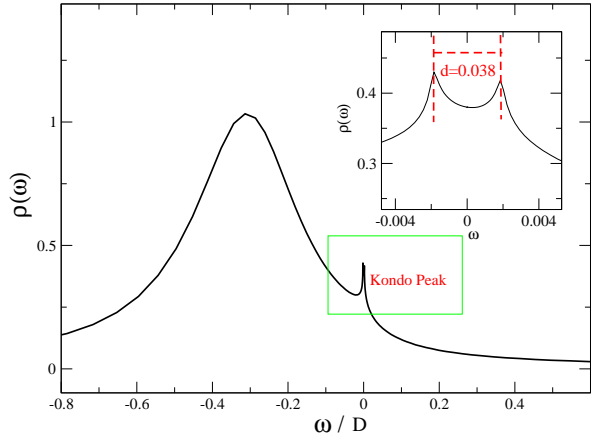}
\end{center}
\par
\vspace{-0.5cm}
\caption{The non-equilibrium impurity spectral function $\rho(\omega)$ 
(in arbitrary unit) with $\Gamma=0.3D$, $\epsilon_d=-0.3 D$, $r=0.05$, $V=0.038 D$, $T=5\times10^{-7}D$. 
The Kondo peak splits into two peaks where the width of the two splited peaks 
equals to bias $V$ (see Inset).}
\vspace{0.5cm}
\label{DOS-bias}
\end{figure}

Here, we employ the NCA to address the equilibrium and non-equilibrium 
transport at criticality based on 
Eq.~(\ref{Hamiltonian}). This approach has been known to correctly capture 
the non-Fermi liquid properties of the 
two-channel Anderson model\cite{Zamani.12,Zamani.13,kondo}. Recently, it 
has been generalized to address the 2CK-LM 
crossover in non-equilibrium transport 
in a voltage-biased 2-channel pseudogap Anderson model with $r=1$, relevant for graphene\cite{NCA-graphene}. We generalize this approach here further to 
the voltage-biased 2-channel pseudogap Anderson model with $0<r<1$. 

Within NCA, the Green's functions for the conduction electrons $G_{\sigma,{\it i}\lambda}(t)$, pseudo-fermions $G_{f\sigma}(t)$, and the slave bosons $D(t)$ are given by\cite{Meir,kroha}: 
\begin{eqnarray}
G^r_{\sigma,i\lambda}(t)&=& -i\theta(t) <\{c_{\sigma}(t),c^\dag_{\sigma}(0) \} >_{i\lambda} \nonumber \\
&=& -i \theta(t)[D^>(-t)G^<_{f\sigma}(t)-D^< (-t) G^>_{f\sigma}(t)].\nonumber \\
\end{eqnarray}
\begin{equation}
G^>_{f\sigma}\equiv -i<f_{\sigma}(t)f^{\dag}_{\sigma}(0)>_{i\lambda}, 
G^<_{f\sigma}\equiv i<f^{\dag}_{\sigma}(0)f_{\sigma}(t)>_{i\lambda},\nonumber \\
\end{equation}
and 
\begin{equation}
D^>\equiv -i<b_\tau (t)b^{\dag}_\tau (0)>_{i\lambda}, 
D^<\equiv i<b^{\dag}_\tau (0) b_\tau (t)>_{i\lambda},\nonumber \\
\end{equation}
where the notation $<$ and $>$ represents lesser and greater Green function. 
The lesser (greater) Green functions can be written as \cite{Meir,kroha}:
\begin{eqnarray}
D^{> (<)}(\omega)&=&D^r(\omega) \Pi^{> (<)}(\omega) D^a(\omega),\nonumber \\
G^{> (<)}_{f\sigma}&=&G^r_{f\sigma}(\omega)\Sigma^{< (>)}_{f\sigma}(\omega) G^a_{f\sigma},
\end{eqnarray}
where $D^{a(r)}(\omega)$ and $G^{a(r)}_{f\sigma}(\omega)$ are advanced (retarded) Green function of boson and fermion, respectively, and the subscript 
${\it i}\lambda$ refers to the enforcement of the local constraint on the 
impurity via Lagrange multiplier $\lambda$ when evaluating these 
correlation functions. 
The $\Pi^{> (<)}(\omega)$ and $\Sigma^{< (>)}_{f\sigma}(\omega)$ are 
the self-energies of slave-boson  and pseudofermion greater (lesser) 
Green functions, respectively. 
The NCA expressions for the lesser self-energy of the pseudofermion, 
$G^{<}_{f\sigma}(\omega)=\Sigma^{<}(\omega)|G_{f\sigma}^{r}(\omega)|^{2}$, and slave-boson, $D^{<}(\omega)=\Pi^{<}(\omega)  |D^{r}(\omega)|^{2}$, are\cite{Meir,kroha,NCA-graphene}:
\begin{figure}[t]
\begin{center}
\includegraphics[width=1.0\linewidth]{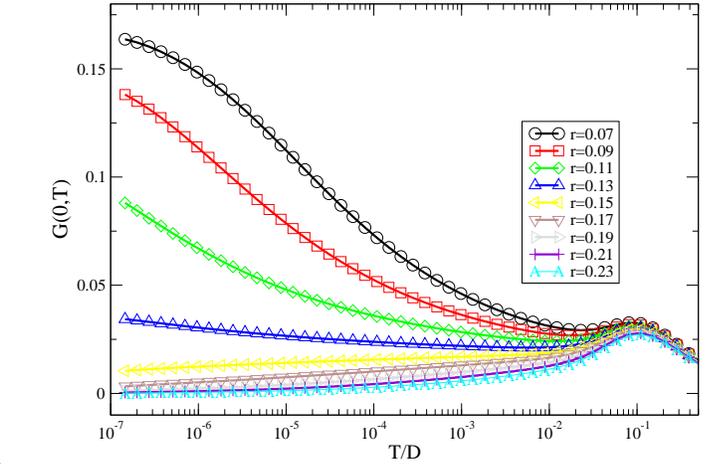}
\end{center}
\par
\vspace{-0.5cm}
\caption{The equilibrium (linear) conductance $G(0,T)$ 
(in unit of $2e/\hbar$) versus temperature $T$ (in unit of $D=1$) 
for various $r$ with $\Gamma=0.3D$, $\epsilon_d=-0.3 D$.}
\vspace{0.5cm}
\label{GT}
\end{figure}

\begin{eqnarray}
\Sigma^{<}_{f\sigma}(\omega)=\frac{2}{\pi}\sum_{\alpha}\int d\epsilon
\Gamma_{\alpha}(\omega-\epsilon-\mu_\alpha)
f(\omega-\epsilon-\mu_\alpha) D^{<}(\epsilon),\nonumber \\
\label{SFeq3}
\end{eqnarray}
\begin{eqnarray}
\Pi^{<}(\omega)=\frac{2}{\pi}\sum_{\alpha}\int d\epsilon
\Gamma_{\alpha}(\epsilon-\omega-\mu_\alpha)
f(\omega-\epsilon+\mu_\alpha) G^{<}_{f\sigma}(\epsilon).\nonumber \\
\label{SFeq4}
\end{eqnarray}
Here, $\Gamma_{\alpha}(\omega)\equiv \Gamma_\alpha \rho_{c,\alpha}(\omega)$ 
with $\Gamma_\alpha= \pi|V_{k\sigma}^\alpha|^{2}$ with
$\rho_{c,\alpha}(\omega)=- \frac{1}{\pi}ImG_{c}(\omega)
=\frac{r+1}{2D^{r+1}}|\omega|^r \theta(D-|\omega-\mu_\alpha|)$ and $D=1$ being 
the bandwidth of the conduction bath, 
and $f(\omega)=\frac{1}{1+e^{\beta \omega}}$ is the Fermi function. 

The relation between greater Green function and retarded Green function are,
$D^{>}(w)=2iImD^{r}(w)$, $G^{>}_{f \sigma}=2iImG^{r}_{f \sigma}$, where the self-energies, $\Pi^>(w)=2iIm\Pi^r(w)$ and $\Sigma^>_{f\sigma}(w)=2iIm\Sigma^r_{f\sigma}(w)$. The NCA expressions for the self energies of retarded Greens functions 
for pseudo-fermion 
$G^r_{f\sigma}(\omega)=[\omega-\epsilon_d- \Sigma^{r}(\omega)]^{-1}$ 
and for slave-boson $D^r(\omega)= [\omega-\Pi^{r}(\omega)]^{-1}$ are given by\cite{Meir,kroha,NCA-graphene}:
\begin{eqnarray}
\Sigma^{r}(\omega)&=&\frac{2}{\pi}\sum_{\alpha}\int d\epsilon
\Gamma_{\alpha}(\omega-\epsilon-\mu_\alpha)f(\epsilon-\omega-\mu_\alpha)D^{r}(\epsilon),\nonumber \\
\label{SFeq1}
\end{eqnarray}
\begin{eqnarray}
\Pi^{r}(\omega)&=&\frac{2}{\pi}\sum_{\alpha}\int d\epsilon
\Gamma_{\alpha}(\epsilon-\omega-\mu_\alpha) 
f(\epsilon-\omega-\mu_\alpha) G^{r}_{f\sigma}(\epsilon).\nonumber \\
\label{SFeq2}
\end{eqnarray}
\begin{figure}[t]
\begin{center}
\includegraphics[width=1.0\linewidth]{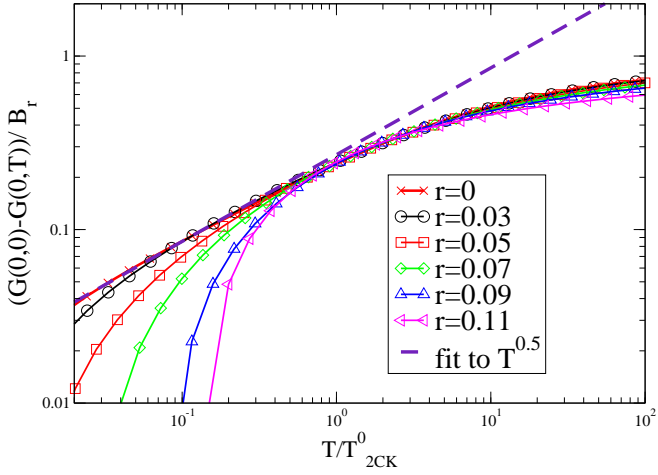}
\end{center}
\par
\vspace{-0.5cm}
\caption{The $
\sqrt{T}$ scaling of equilibrium (linear) conductance $(G(0,0)-G(0,T))/B_r$ 
(in unit of $2e/\hbar$) versus $T/T_{2CK}^0$ for various $r$ with $\Gamma=0.28D$, $\epsilon_d=-0.2 D$ with $T_{2CK}^0$ being the two-channel Kondo temperature 
for $r=0$. 
Here, $B_r$ are non-universal functions of $r$. 
The dashed line is a fit to $\sqrt{T}$ behavior.}
\vspace{0.5cm}
\label{GT_2CK}
\end{figure}
The physical impurity spectral function, $\rho_{\sigma}(\omega,V)$, is obtained via 
the convolution of pseudo-fermion and slave-boson Greens function based on 
Eqs.~(\ref{SFeq3}),(\ref{SFeq4}),(\ref{SFeq1}), and (\ref{SFeq2}) as\cite{NCA-graphene}: 
\begin{eqnarray}
\rho(\omega,V)=\frac{i}{2\pi^{2}Z}\int d\epsilon[ImD^{r}(\epsilon)G^{<}(\omega+\epsilon)-\nonumber\\D^{<}(\epsilon)ImG^{r}(\omega+\epsilon)].
\label{DOSeq}
\end{eqnarray}
The normalization factor $Z=\frac{i}{2\pi}\int d\omega [M\times D^<(\omega)-N\times G^<(\omega)]$ enforces the constraint, $<Q>=1$ with $M=N=2$ here.
The current going through the impurity therefore reads\cite{NCA-graphene}:
\begin{eqnarray}
I(V,T)&=&\frac{2e}{\hbar}\int d \omega \frac{2\Gamma_{L}(\omega)\Gamma_{R}(\omega)}
{\Gamma_{L}(\omega)+\Gamma_{R}(\omega)} \rho(\omega,V,T)\nonumber \\
&\times& [f(\omega+eV/2)-f(\omega-eV/2)].
\label{IVeq}
\end{eqnarray}
where $\Gamma_\alpha(\omega) = \Gamma(\omega-\mu_\alpha)$ with $\alpha=L,R$. 
The equilibrium (linear) conductance is directly obtained via 
\begin{eqnarray}
G(0,T)&=&\frac{2e^{2}}{\hbar}\int d\omega 
\frac{2\Gamma_{L}(\omega)\Gamma_{R}(\omega)}
{\Gamma_{L}(\omega)+\Gamma_{R}(\omega)}
\Big(-\frac{\partial f(\omega)}{\partial\omega}\Big)\nonumber \\
&\times& \rho(\omega,V=0).
\label{GTeq}
\end{eqnarray}
And the nonlinear conductance $G(V)$ is given by $\frac{dI(V)}{dV}$.
Eqs.(6)-(10) form a self-consistent set of Dyson's equations within NCA. 
We solve these equations self-consistently and evaluate  
Eqs.(\ref{DOSeq})-(\ref{GTeq}) based on these solutions.

\begin{figure}[t]
\begin{center}
\includegraphics[width=\linewidth]{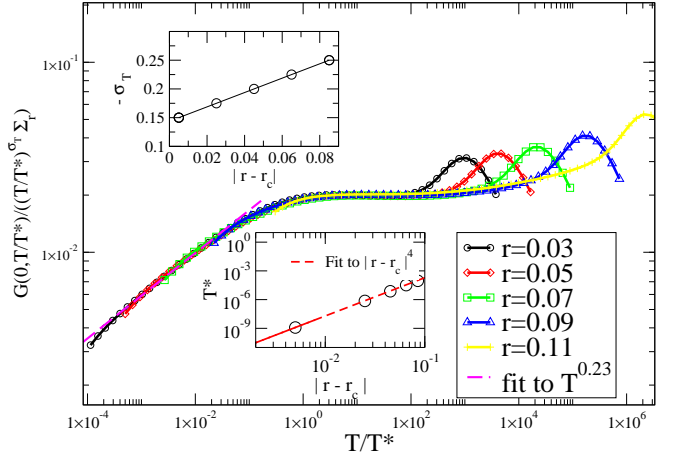}
\end{center}
\par
\vspace{0.0cm}
\caption{Universal scaling in linear conductance $G(0,T/T^\ast)$ 
(in unit of $2e/\hbar$ and normalized to $\Sigma_r (T/T^\ast)^{\sigma_T}$) as a function 
of temperature $T/T^\ast$ near 2CL-LM quantum phase transition 
for various values of $r$. Parameters are the same as in Fig.~\ref{GT}.
Here, $T^\ast$, $\sigma_T$, and $\Sigma_r$ are defined in the text. The 
magenta dashed line is a fit to $T^{0.23}$. Top Inset: 
The power-law exponent $\sigma_T$ in linear conductance 
$G_{QCP}(T)$ close to criticality as a function of $|r-r_c|$. 
Solid line is a fit to a linear relation: $\sigma_T = \beta_T -\alpha_T |r-r_c|$ with $\beta_T\approx -0.145$, $\alpha_T\approx 1.25$. Bottom Inset: Crossover temperature $T^\ast$ versus $|r-r_c|$. The red dashed line is a fit to $|r-r_c|^4$.}
\vspace{0.5cm}
\label{GT-scaling}
\end{figure}

\section{\bf III. Results.} 

\subsection{A. Quantum critical point at 2CK-LM phase transition: impurity spectral function}

The existence of a quantum critical point (QCP) separating 
2CK for $r<r_c$ from the LM for $r>r_c$ 
phases exists in the PSG Anderson model 
has been studied extensively\cite{Vojta-2CKPSG,Zamani.12}. 
The generic phase diagram is 
shown as in Fig.~\ref{phase}.  
We focus here on the transport properties for the two-lead 
setup near criticality both in and out of equilibrium, especially 
on the distinct non-equilibrium quantum critical properties 
(see Fig.~\ref{phase}). 
In equilibrium ($\mu_\alpha=0$), 
the 2CK-LM phase transition is studied here by tuning 
$r$ of the pseudogap power-law DOS of the leads with fixed hybridization 
parameter $\Gamma_\alpha$ and the impurity energy $\epsilon_d$. 
The value of $r_c$ is extracted from the local impurity spectral 
function $\rho_{\sigma}(\omega,V=0)$ 
via solving the self-consistent Dyson's equations. 
Since $\epsilon_d\neq -U/2$, the 2CK pseudogap Anderson model considered here 
shows particle-hole (ph) asymmetry, giving rise to an overall 
asymmetric impurity spectral function with respect to the Fermi energy 
(see Fig.~\ref{DOSfig}). 
For $r<r_c$, $\rho_\sigma(\omega)$ shows a non-Lorentzian 
Kondo peak, a characteristic of the non-Fermi liquid 2CK 
state\cite{kondo,kroha}. 
In fact, $\rho_\sigma(\omega)$ exhibits a power-law singularity 
near $\omega=0$ with an exponent being $r$: 
$\rho_\sigma(\omega)\sim \omega^{-r}$\cite{Zamani.12}. 
With increasing $r$, the Kondo peak gets narrower with reduced 
spectral weight.  
For $r\le r_c$, however, the Kondo peak splits into two, and the ground state 
is in the LM phase. The critical value $r_c\approx 0.115$ for 
$\Gamma_\alpha \sim 0.3D$, $\epsilon_d\sim -0.2D$ (see Fig.~\ref{DOSfig}). 
The spectral weight of the 
Kondo peaks gets further suppressed with increasing $r$ until it completely 
disappears. 
At a finite bias voltage for $r<r_c$, 
the impurity local DOS shows 
splited Kondo peaks at $\omega=\pm V/2$ (see Fig.~\ref{DOS-bias}). 
Note that the non-zero 
local DOS of $\rho_\sigma(\omega=0)$ for $r>r_c$ 
is due to  
the limitation of the lowest temperature $T_0~\sim 5\times 10^{-7}D$ we 
can reach numerically. As $T\rightarrow 0$ and for $r\ge r_c$, 
we expect a complete dip developed 
in local DOS such that $\rho_\sigma(\omega=0)\sim 0$.  
 
\begin{figure}[t]
\begin{center}
\includegraphics[width=1.0\linewidth]{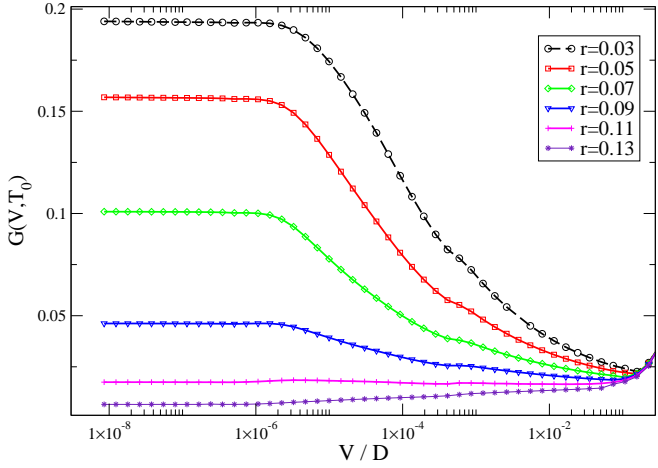}
\end{center}
\par
\vspace{-0.5cm}
\caption{The non-equilibrium (non-linear) conductance $G(V,T_0)$ (in unit of $2e/\hbar$) versus bias voltage $V$ (in unit of $D=1$) 
for various $r$ with $\Gamma=0.3D$, $\epsilon_d=-0.3 D$.}
\vspace{0.5cm}
\label{GV}
\end{figure}

\subsection{B. Universal scaling in linear (equilibrium) conductance near criticality}

More clear signatures of the 2CK-LM quantum phase transition can be 
obtained via linear and non-linear conductances. First, we analyze the 
linear (equilibrium) conductance at finite temperatures but zero bias voltage 
$G(V=0,T)$. Fig.~\ref{GT} shows $G(0,T)$ for different $r$ 
with $\Gamma=0.28D$, $\mu_0=0$, $\epsilon_d=-0.2D$, and $D=1$. 
For $r=0$ it is well known that $G(0,T)$ follows the 2CK scaling 
function\cite{kroha}: $G(0,T)-G(0,0)=B_r \sqrt{T/T_{2CK}^0}$ for $T<T_{2CK}^0$ 
with $T_{2CK}^0\sim 3\times 10^{-5}D$ being the 2CK Kondo temperature at $r=0$ 
and $B_r$ being a non-universal constant. 
Here, we set $G(0,0)\simeq G(0,T_0=5\times10^{-7}D)$. 
For $0<r<r_c$, however, the 2CK $\sqrt{T}$ scaling in $G(0,0)-G(0,T)$ ceases to 
exist. As shown in Fig.~\ref{GT}, 
the $\sqrt{T}$ behavior in $G(0,0)-G(0,T)$ is clear for $r=0$, 
but it deviates from $\sqrt{T}$ more with increasing $r<r_c$. 
We propose that this deviation from the conventional 
2CK behavior for $T<T_{2CK}$ 
for $0<r<r_c$ 
can be due to the following two scenarios: (i). 
the emergence of distinct universal power-law 
scaling behaviors when  
the system approaches the 2CK-LM quantum critical regime for $T>T^\ast$ 
with $T^\ast$ being 
the crossover energy scale above which quantum critical behaviors 
are observed,  
and (ii). the existence of a distinct 2CK scaling form other 
than $\sqrt{T}$ in conductance 
in low temperature regime $T<T_{2CK}$ for $0<r<r_c$ with $T_{2CK}$ being 
the two-channel Kondo temperature for $0<r<r_c$. 
As indicated in Fig.~\ref{GT_2CK}, 
the generic 2CK behaviors for $0<r<r_c$ can not be manifested in 
$G(0,0)-G(0,T)$ as an universal power-law in $T$ as it does for $r=0$ case.
Therefore, instead of analyzing $G(0,0)-G(0,T)$, we therefore address the 
above  two scenarios below via trying the possible 
scaling behaviors of $G(T)$ itself.  
\begin{figure}[t]
\begin{center}
\includegraphics[width=1.0\linewidth]{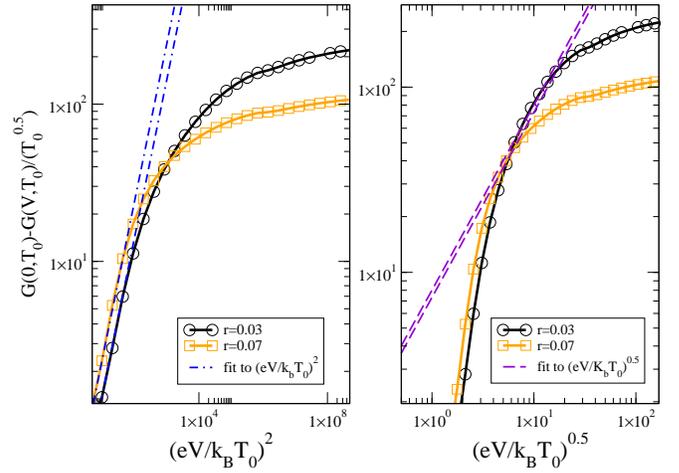}
\end{center}
\par
\vspace{-0.5cm}
\caption{The 2CK scaling of the 
non-equilibrium (non-linear) conductance $(G(0,T_0)-G(V,T_0))/T_0^{0.5}$ 
(in unit of $2e/\hbar$) versus (a) $(eV/k_BT)^2$ and (b) $(eV/k_B T)^{0.5}$ 
for various $r$ with $\Gamma=0.3D$, $\epsilon_d=-0.3 D$.}
\vspace{0.5cm}
\label{GV_sqsqrt}
\end{figure}

First, as $r\rightarrow r_c$, the existence of a quantum critical regime 
for $T>T^\ast$
requires 
the divergence of the correlation length $\xi$ in a power-law fashion: 
$\xi\propto |r-r_c|^{-\nu}\rightarrow\infty$\cite{QPT,vojta-review}. 
As a result,  
all thermaldynamical observables, including conductances,  
are expected to exhibit universal scaling properties. 
As shown in Fig.~\ref{GT-scaling}, 
we indeed find numerically that the linear conductance $G_{QCP}(0,T)$ 
shows an universal power-law in $T$ near criticality  within a temperature 
range of (approximately) $5\times10^{-7}D <T <5\times10^{-4}D$ as:
\begin{equation}
G_{QCP}(0,T) \propto T^{\sigma_T}=T^{\beta_T -\alpha_T |r-r_c|},
\end{equation}
where the exponent $\sigma_T$ exhibits a linear relation to 
$|r-r_c|$ with $\beta_T\approx -0.145$ and $\alpha_T\approx 1.25$ 
being non-universal constant pre-factors dependent on $\Gamma$ and 
$\epsilon_d$, and  
$T^{\sigma_T}=T^{\beta_T}$ as $r=r_c\sim 0.115$. 
Based on the above analysis, we divide 
$G(0,T)$ by the power-law function $T^{\sigma_T}$, gives:
\begin{eqnarray}
\widetilde{G}(0,T)&=& G(0,T)/T^{\sigma_T}=\frac{G(0,T)}{T^{\beta_T -\alpha_T |r-r_c|}}, \nonumber \\
\widetilde{G}_{QCP}(0,T)&=&G_{QCP}(0,T)/T^{\sigma_T}
\end{eqnarray}
where $\widetilde{G}_{QCP}(0,T)$ in the quantum critical region becomes a 
constant: $\widetilde{G}_{QCP}(0,T)\sim G_{QCP}^0$. 
The universal scaling function $\overline{G}(0,\frac{T}{T^*})$ 
in linear conductance is obtained 
by rescaling $\widetilde{G}(0,T)$ by a non-universal factor $\Sigma_r$ and  
$T$ by the crossover energy scale $T^\ast$ (see Fig.~\ref{GT-scaling}):
\begin{equation}
\overline{G}(0,\frac{T}{T^*})\equiv \frac{\tilde{G}(0,\frac{T}{T^*})/\Sigma_r}{(T/T^*)^{\beta_T -\alpha_T |r-r_c|}},
\end{equation}
where $T^*$ is proportional to the inverse of 
correlation length $1/\xi_T \propto |r-r_c|^{\nu_T}$, vanishing in a power-law 
of $|r-r_c|$ with the exponent $\nu_T$ being the correlation length exponent:
\begin{equation}
T^* \propto \frac{1}{\xi_T} = D |r-r_c|^{\nu_T}
\end{equation}
with $\nu_T$ being the correlation length exponent 
corresponding to the power-law exponent of crossover scale 
$T^*$ versus $|r-r_c|$. We find $\nu_T\sim 4$ here.  
(see Inset of Fig.~\ref{GT-scaling}).
 As $r\rightarrow r_c$ from below, 
we find a perfect data collapse in 
$\overline{G}(0,\frac{T}{T^*})\approx const.$ 
for $1<T/T^\ast < 10^3$, indicating 
the range of quantum critical region. 
Surprisingly, at low temperatures $T<T_{2CK}\approx 0.1 T^\ast$ 
where the system is governed by the 2CK phase, we find 
the above function 
$\overline{G}(0,\frac{T}{T^*})$ exhibits as a distinct universal 
function power-law scaling function (see Fig.~\ref{GT-scaling}): 
\begin{eqnarray}
\overline{G}(0,\frac{T}{T^*})_{2CK}&\propto& 
(\frac{T}{T_{2CK}})^{\sigma_{2CK}^T} 
\label{GT-2CK}
\end{eqnarray}
with $\sigma_{2CK}^T\sim 0.23$ 
for the parameters we set 
(or equivalently $G(\frac{T}{T_{2CK}})\propto (\frac{T}{T_{2CK}})^{\sigma_{2CK}^T+\sigma_T}$). 
We may therefore regard this low temperature behavior  
in $G(0,T)$ as the distinct 2CK scaling in equilibrium (linear) 
conductance for the pseudogap 2-channel Anderson model.

\begin{figure}[t]
\begin{center}
\includegraphics[width=\linewidth]{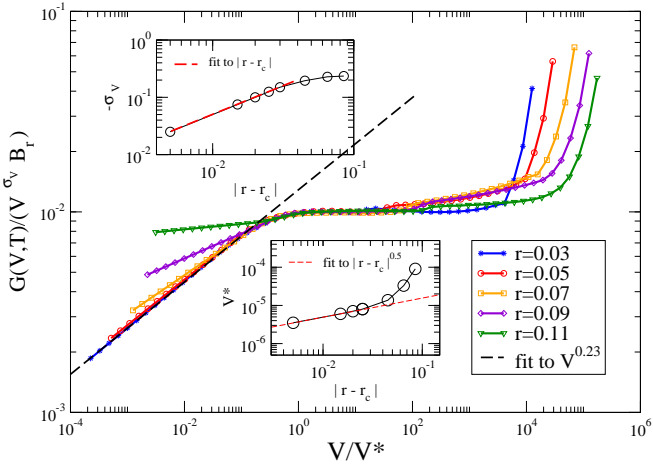}
\end{center}
\par
\vspace{0.0cm}
\caption{Universal scaling in non-linear conductance $G(V,T_0)/(V^\sigma_V B_r)$ (in unit of $2e/\hbar$) as a function 
of temperature $V/V^\ast$ near 2CL-LM quantum phase transition 
for various values of $r$. Parameters are the same as in Fig.~\ref{GT}.
Here, $V^\ast$, $\sigma_T$, and $B_r$ are defined in the text. Black dashed line is a power-law fit to the conductance in the 2CK regime. Top Inset: 
The power-law exponent $\sigma_V$ in non-linear conductance 
$G_{QCP}(V,T_0)$ close to criticality as a function of $|r-r_c|$. 
Solid line is a fit to a linear relation: $\sigma_V = \beta_V -\alpha_V |r-r_c|$ with $\beta_V\approx 0$, $\alpha_V\approx 2.5$. Bottom Inset: Crossover temperature $V^\ast$ versus $|r-r_c|$. The red dashed line is a fit to $|r-r_c|^{0.5}$.}
\vspace{0.5cm}
\label{GV-scaled}
\end{figure}

\subsection{B. Universal scaling in non-linear (non-equilibrium) conductance near criticality}

We now add the bias voltage in the leads and study the scalings in 
non-equilibrium conductance near 2CK-LM quantum critical point. 
It is generally expected that the scaling behaviors of non-linear conductance 
near criticality are distinct from that in equilibrium\cite{QPT-noneq}.  
The non-linear conductance 
$G(V,T_0)$ is obtained 
at a fixed low temperature $T_0=5\times 10^{-7}$, symmetrical 
hybrizidation $\Gamma_{L}=\Gamma_R=0.3D$, and $\epsilon_d=-0.3D$. As 
shown in Fig.~\ref{GV}, 
the non-linear conductance $G(V,T_0)$ for each value of $r$ 
saturates as $V\rightarrow 0$, while a small ``kink''-like minima in 
conductance is developed at $eV\approx k_B T$. 
For $r=0$ and $T<T_{2CK}^0$, it follows the 
well-known 2CK scaling\cite{kroha}: 
$G(0,T_0)-G(V,T_0) = \Sigma_rT^{\frac{1}{2}}H(\frac{eV}{k_B T_0})$, 
where the universal function $H(\frac{eV}{k_BT_0})$ behaves as 
$(\frac{eV}{k_BT_0})^2$ for $eV \le k_BT_0$, and 
$(\frac{eV}{k_BT_0})^{\frac{1}{2}}$ 
for $eV\gg k_B T_0$. 
However, for $0<r<r_c$ 
as the system gets closer to criticality, while 
$(\frac{eV}{k_BT_0})^2$ behavior remains in $G(0,T_0) - G(V,T_0)$ 
for $eV\sim k_BT_0$, more deviations from the 2CK 
$(\frac{eV}{k_BT_0})^{\frac{1}{2}}$ behavior 
are observed for $k_BT_0<eV<k_B T_{2CK}$. 
Furthermore, similar to the case in our analysis for equilibrium conductance, 
as indicated in Fig.~\ref{GV_sqsqrt}, 
the universal 2CK behavior in non-linear conductance 
for $0<r<r_c$ in general are not manifested in $G(0,T_0) - G(V,T_0)$ in 
the power-law fashion as it is for $r=0$ case. 
We therefore analyze the universal scaling properties based on 
$G(V,T_0)$ itself rather than $G(0,T_0) - G(V,T_0)$. 
We perform the similar analysis here to that in equilibrium. 
As $r\rightarrow r_c$, we find $G(V,T_0)$ shows a power-law dependence on $V$ 
in the quantum critical regime approximately $10^{-4}< V/D < 10^{-6}$, defined 
as $G_{QCP}(V,T_0)$ with power-law exponent linear in $|r-r_c|$:
\begin{equation}
G_{QCP}(V,T_0)\propto V^{\sigma_V} = V^{\beta_V-\alpha_V|r-r_c|}
\label{GV-QCP}
\end{equation}
with $\beta_V$, $\alpha_V$ 
being non-universal prefactors 
dependent on $\Gamma$ and $\epsilon_d$. At criticality $r=r_c$, $\sigma_V\sim 0$, yielding a constant non-linear conductance: 
$G_{QCP}(V,T_0)|_{r_c}\sim const.$. We further perform a re-scaling on 
$G(V,T_0)$ as:
\begin{eqnarray}
\tilde{G}(V,T_0) &\equiv& \frac{G(V,T_0)}{V^{\sigma_V}} = 
\frac{G(V,T_0)}{V^{\beta_V-\alpha_V|r-r_c|}}
\label{tilde-GV}
\end{eqnarray}
where $\tilde{G}(V,T_0)$ becomes a constant in the critical regime. 
We may further define the universal scaling function 
$\bar{G}(\frac{V}{V^\ast},T_0)$ for non-linear conductance via 
the following re-scalings: 
$V\rightarrow V/V^\ast$, $\tilde{G}(V,T_0)\rightarrow 
\tilde{G}(V,T_0)/\Sigma_r$ (see Fig.~\ref{GV-scaled}):
\begin{eqnarray}
\bar{G}(\frac{V}{V^\ast},T_0)&=&
\frac{G(\frac{V}{V^\ast},T_0)/\Sigma_r}{(\frac{V}{V^\ast})^{\beta_V-\alpha_V|r-r_c|} }
\label{GV-scaling}
\end{eqnarray}
 where $V^\ast$ is the crossover energy scale, and 
$\Sigma_r$ a non-universal constant pre-factor. Here, 
we find the exponent $\sigma_V$ depends linearly on $|r-r_c|$ with  
$\beta_V\approx 0$, $\alpha_V\approx 2.5$ by the best fit of the data 
(see Inset of Fig.~\ref{GV-scaled}), and  
the  crossover scale $V^\ast$ is inversely 
proportional to the correlation length $\xi_V$ with a power-law dependence 
on the distance to criticality: $V^\ast\propto \frac{1}{\xi_V}\propto |r-r_c|^{\nu_V}$ with $\nu_V\sim 0.5$ (see Inset of Fig.\ref{GV-scaled}). 
Note that these critical 
exponents out of equilibrium are distinct from those in equilibrium, 
and can be considered as characteristics of non-equilibrium quantum 
criticality. The distinction between equilibrium and nonequilibrium 
quantum critical properties is expected due to the different role 
played by the temperature and voltage bias near criticality, leading to 
different behaviors in decoherence rate $\Gamma^s$ 
(the broadening of impurity DOS) in equilibrium 
$\Gamma_T^s$ versus out of equilibrium $\Gamma_V^s$\cite{QPT-noneq,chung-2ckpsg}. 
 
For $T_0<V< V^\ast$, the system approaches 2CK state at a characteristic 
energy scale $V\sim V_{2CK}\approx 0.1 V^\ast$ below which an universal 
power-law scaling is observed (see Fig.~\ref{GV-scaled}):
\begin{eqnarray}
\bar{G}(\frac{V}{V^\ast},T_0)_{2CK}&\propto& (\frac{V}{V\ast})^{\sigma^V_{2CK}}
\label{GV-2CK}
\end{eqnarray}
where the exponent $\sigma_{2CK}^V\sim 0.23$. 
In the low bias limit $V<T_0$, 
conductance reaches a constant equilibrium 
value $G(0,T_0)$ and 
therefore deviates from the 2CK scaling (see Fig.~\ref{GV-scaled}).

Note that we find 
Eq.~(\ref{GV-2CK}) for non-equilibrium conductance 
to exhibit the same form as that  
in equilibrium in Eq.~(\ref{GT-2CK}) with the same exponent 
($\sigma^T_{2CK}=\sigma_{2CK}^V\sim 0.23$). 
Further studies are required to clarify if this 
relation holds in general. 
Though the exact value ($0.23$) of the above exponents 
depend in general on physical parameters, such as: 
$\epsilon_d$ and $\Gamma$,  
Eqs.~(\ref{GT-2CK}) 
and (\ref{GV-2CK}) correctly reproduce  
the well-known $\sqrt{V}$ and $\sqrt{T}$ behaviors for $r=0$ 
in the 2CK regime in $G(0,T_0)-G(V,T_0)$ and $G(0,T_0) - G(0,T)$, 
respectively. 

\section{\bf V. Conclusions.} 

In conclusions, we have studied quantum phase transitions in and out 
of equilibrium between two-channel 
Kondo and local moment phases in the two-channel pseudogap Anderson model 
where the conduction leads show a power-law vanishing density of states 
with exponent $0<r<1$. 
Via Non-Crossing Approximation (NCA), we solved self-consistently for 
the impurity Green's function, and therefore determined the linear and 
non-linear conductances. The 2CK-LM quantum criticality is reached 
by varying the power-law exponent $r$ of the pseudogap conduction bath 
with fixed lead-dot hybridizations and chemical potentials. 
The linear $G(V=0,T)$ and non-linear $G(V,T_0)$ ($T_0\sim5\times10^{-7}D$) 
conductances show distinct universal 
power-law scalings near criticality  
with different critical exponents. Furthermore, in the 2CK regime 
$T,V<T_{2CK}$, we also find different 
characteristic power-law scalings in $G(T)$ and $G(V,T_0)$ compared to the 
well-known $\sqrt{T},\sqrt{V}$ scalings for $G(0,0) - G(0,T)$ 
and $G(V,0)-G(V,T_0)$, respectively. 
Our results provide further insights on 
two-channel Kondo physics and on 
the non-equilibrium quantum criticality in nano-systems. Further 
analytic and numerical investigations are needed to 
address issues on the mechanism behind the different 
scalings between equilibrium and nonequilibrium conductances.

\acknowledgments
We thank S. Kirchner and K. Ingersent for useful discussions. C.H. C. acknowledges the NSC grant No.98-2112-M-009-010-MY3, No.101-2628-M-009-001-MY3, the NCTS, the MOE-ATU program of Taiwan R.O.C.


\begin{thebibliography}{36}

\bibitem{QPT}
S. Sachdev, Quantum phase transitions, Cambridge University press (2000); 
S. L. Sondhi, S. M. Girvin, J. P.
Carini, and D. Shahar, Rev. Mod. Phys. {\bf 69}, 315 (1987).

\bibitem{vojta-review}
M. Vojta, Phil. Mag. {\bf 86}, 1807 (2006).

\bibitem{QD}
L. Kouwenhoven and L. Glazman, Physics World {\bf 14}, 33 (2001); 
D. Goldhaber-Gordon et al., Nature {\bf 391}, 156 (1998);
W. G. van der Wiel et al., Science {\bf 289}, 2105 (2000); 
L.I. Glazman and M.E. Raikh, Sov. Phys. JETP Lett. {\bf 47}, 452 (1988); 
T.K. Ng, P.A. Lee, Phys. Rev. Lett. {\bf 61}, 1768 (1988).

\bibitem{kondo}
A.C. Hewson, The Kondo Problem to Heavy Fermions (Cambridge University 
Press, Cambridge, UK, 1997).

\bibitem{QPT-noneq}
Chung-Hou Chung, Karyn Le Hur, Matthias Vojta, Peter Wölfle, 
Phys. Rev. Lett. 102, 216803 (2009);  
Chung-Hou Chung, Karyn Le Hur, Gleb Finkelstein, Matthias Vojta, Peter Woelfle,
 Phys. Rev. B {\bf 87}, 245310 (2013); 
Chung-Hou Chung, Kenneth Yi-Jie Zhang, 
Phys. Rev. B {\bf 85}, 195106 (2012);S. Kirchner and Q. Si, Phys. Rev. Lett. {\bf 103}, 206401 (2009). 

\bibitem{2CK}
P. Nozieres, A. Blandin, J. Phys. (Paris) {\bf 41}, 193 (1980).

\bibitem{Affleck-Ludwig}
I. Affleck, and A.W.W. Ludwig, Nucl. Phys. B {\bf 360}, 641 (1990); 
I. Affleck and A.W.W. Ludwig, Phys. Rev. Lett. {\bf 67}, 161 (1991).

\bibitem{kroha}
M. H. Hettler, J. Kroha, and S. Hershfield, Phys. Rev. Lett. {\bf 73}, 1968 (1994); M.H. Hettler, J. Kroha, and S. Hershfield, Phys. Rev. B {\bf 58}, 
5649 (1998). 


\bibitem{Fendley}
P. Fendley, F. Lesage, and H. Saleur, J. Stat. 
Phys. {\bf 79}, Nos. 5/6, 799 (1995).


\bibitem{nayak}
Paul Fendley, Matthew P.A. Fisher, Chetan Nayak, Phys. Rev. B {\bf 75}, 
045317 (2007); Gregory A. Fiete, Waheb Bishara, Chetan Nayak, Phys. Rev. 
Lett. {\bf 101}, 176801 (2008).

\bibitem{sela} 
Eran Sela, Andrew K. Mitchell, Lars Fritz, Phys. Rev. Lett. {\bf 106}, 
147202 (2011); Andrew K. Mitchell, Eran Sela, and David E. Logan, Phys. Rev. Lett. {\bf 108}, 086405 (2012).


\bibitem{Bethe}
N. Andrei and C. Destri, Phys. Rev. Lett. {\bf 52}, 364 (1984), 
 P. B. Wiegmann and A. M. Tsvelik, Z. Phys. B {\bf 54}, 201 (1985).

\bibitem{CFT}
I. Affleck and A. W. W. Ludwig, Nucl. Phys. B {\bf 352}, 849
(1991); ibid. {\bf 360}, 641 (1991). A.W.W. Ludwig, I. Affleck, 
Phys. Rev. Lett. {\bf 67}, 3160 (1991), E. Sela, A.K. Mitchell, L. Fritz, Phys. Rev. Lett. {\bf 106}, 147202 (2011).

\bibitem{bosonization}
V.J. Emery, S. Kivelson, Phys. Rev. B {\bf 46}, 10812 (1992); M. Fabrizio, A.O. Gogolin, Phys. Rev. B {\bf 51}, 17827 (1995).

\bibitem{2CK-NRG}
A. K. Mitchell, E. Sela, D. E. Logan, Phys. Rev. Lett. {\bf 108}, 086405 (2012).

\bibitem{Goldhaber}
R.M. Potok, I.G. Rau, H. Shtrikman, Y. Oreg, and Goldhaber-Gordon, 
Nature London {\bf 446}, 167 (2007).

\bibitem{Cox_a}
D. L. Cox and A. Zawadowski,Adv. Phys. {\bf 47}, 599 (1998). 

\bibitem{Cichorek}
T. Chichorek et al., Phys. Rev. Lett. {\bf 94}, 236603 (2005).

\bibitem{sengupta}
K. Sengupta and G. Baskaran, Phys. Rev. B {\bf 77}, 045417 (2008).

\bibitem{Vojta-2}
Matthias Vojta, Lars Fritz and Ralf Bulla, Europ. Phys. Lett. {\bf 90}, 27006 (2010).

\bibitem{Mattos}
L. S. Mattos {\it et al.}, (un-published); 
L.S. Mattos, ``Correlated electrons probed by scanning tunneling 
microscopy'', PhD. thesis, Stanford University (2009).

\bibitem{Fradkin}
David Withoff and Eduardo Fradkin, Phys. Rev. Lett. {\bf 64}, 1835 (1990).

\bibitem{Gonzalez-Buxton.98}
C. Gonzalez-Buxton and K. Ingersent, Phys. Rev. B {\bf 57}, 14254 (1998).

\bibitem{Ingersent}
K. Ingersent and Q. Si, Phys. Rev. Lett. {\bf 89}, 076403 (2002).

\bibitem{Glossop_a}
Matthew Glossop et al., Phys. Rev. Lett. {\bf 107}, 076404 (2011).


\bibitem{Vojta4}
Lars Fritz and Matthias Vojta, Phys. Rev. B {\bf 70}, 214427 (2004).


\bibitem{Vojta3}
I. Schneider, L. Fritz, F. B. Anders, A. Benlagra, M. Vojta, 
Phys. Rev. B {\bf 84}, 125139 (2011).

\bibitem{Zamani.12}
F.\ Zamani, T. \ Chowdhury, P. \ Ribeiro, K.\ Ingersent, and S.\ Kirchner, pss in print (2013) and  arXiv:1211.4450.

\bibitem{Zamani.13}
S. Kirchner, F.\ Zamani, E. Munoz, chapter contributed to ``New Materials for Thermoelectric Applications: Theory and Experiment'', Springer Series: NATO Science for Peace and Security Series - B: Physics and Biophysics, Veljko Zlatic (Editor), Alex Hewson (Editor). ISBN: 978-9400749863 (2012), arXiv:1301.3307.

\bibitem{NCA-graphene}
Tsung-Han Lee, Kenneth Yi-Jie Zhang, Chung-Hou Chung, Stefan Kirchner, 
Phys. Rev. B {\bf 88}, 085431 (2013).

\bibitem{Cox_b}
D. L. Cox and A. L. Ruckenstein,  Phys. Rev. Lett. {\bf 71}, 1613 (1993).

\bibitem{Meir} 
Y. Meir and Ned S. Wingreen, Phys. Rev. Lett. {\bf 68}, 2512 (1992); 
Ned S. Wingreen and Y. Meir, Phys. Rev. B {\bf 49}, 11040 (1994).



\bibitem{vojta-largeN}
M. Vojta, Phys. Rev. Lett.  {\bf 87}, 097202 (2001).

\bibitem{Wu-TP}
Tsan-Pei Wu, Mater Thesis, Department of Electrophysics, National Chiao-Tung University, 2014.

\bibitem{Vojta-2CKPSG}
Imke Schneider, Lars Fritz, Frithjof B. Anders, Adel Benlagra, Matthias Vojta, Phys. Rev. B {\bf 84}, 125139 (2011).


\bibitem{chung-2ckpsg}
C.H. Chung {\it et al.}, in preparation.

















 


















\end{thebibliography}
\end{document}